\documentclass[twocolumn,showpacs,preprintnumbers,amsmath,amssymb,superscriptaddress,aps,prc]{revtex4-1}
\usepackage{graphicx}
\usepackage{xcolor}
\usepackage{subfigure}
\def\scr{\rm\scriptscriptstyle }
\usepackage{amssymb}
\usepackage{graphicx}
\usepackage{subfigure}
\usepackage{amsmath}
\usepackage[mathscr]{eucal}

\def\scr{\rm\scriptscriptstyle }

\begin{document}

\title{	Quantum theory of complete and incomplete fusion in collisions of weakly bound nuclei}

\author{J. Rangel} 
\email{jeannie@if.uff.br}
	\affiliation{Instituto de F\'{\i}sica, Universidade Federal Fluminense, Av. Litoranea s/n, Gragoat\'{a}, Niter\'{o}i, R.J., 24210-340, Brazil}

\author{M. R. Cortes} 
	\email{mariane.cortess@gmail.com}	
	\affiliation{Instituto de F\'{\i}sica, Universidade Federal Fluminense, Av. Litoranea s/n, Gragoat\'{a}, Niter\'{o}i, R.J., 24210-340, Brazil}	

\author{J. Lubian} 
	\email{lubian@if.uff.br}
	\affiliation{Instituto de F\'{\i}sica, Universidade Federal Fluminense, Av. Litoranea s/n, Gragoat\'{a}, Niter\'{o}i, R.J., 24210-340, Brazil}

\author{L.F. Canto}
	\email{canto@if.ufrj.br}
	\affiliation{Instituto de F\'{\i}sica, Universidade Federal do Rio de Janeiro, CP 68528, 21941-972, Rio de Janeiro, RJ, Brazil}

\begin{abstract}
We propose a new quantum mechanical method to evaluate complete and incomplete fusion in collisions of weakly bound nuclei. The method is
applied to the $^{7}{\rm Li}+^{209}{\rm Bi}$ system and the results are compared to experimental data. The overall agreement between theory
and experiment is very good, above and below the Coulomb barrier.
\end{abstract}

\maketitle

Nuclear reactions with weakly bound projectiles is one of the main research topics in low energy Nuclear Physics~\cite{CGD06,CGD15,KRA07,KAK09,KGA16,LiS05}. 
Owing to the low breakup threshold, all reaction channels are influenced by the continuum, and
fusion is particularly affected. Besides direct complete fusion (DCF), where the whole projectile merges with the target, there are fusion 
processes following breakup. There is incomplete fusion (ICF), when only non of the breakup fragments fuses with the target, and  
sequential complete fusion (SCF), when all fragments are sequentially absorbed by the target. The sum ${\rm DCF+SCF}$ is called
complete fusion (CF),  and the sum of all fusion processes, that is ${\rm DCF+SCF+ICF}$, is called total fusion (TF).\\

Determining CF and ICF cross sections has been a great challenge for both experimentalists and theorists. Most experiments 
determine only TF cross sections. However, individual CF and ICF measurements have been performed for a few particular projectile-target 
combinations. Dasgupta {\it et al.} measured CF and ICF cross sections in collisions of $^{6,7}$Li projectiles on 
$^{209}{\rm Bi}$~\cite{DHH02,DGH04} and of $^9$Be projectiles on $^{208}$Pb~\cite{DHB99,DGH04}. Similar experiments have been performed to study collisions 
of $^{6,7}$Li projectiles on $^{159}$Tb~\cite{MRP06,BIH75,PMB11}, $^{144,152}$Sm~\cite{RSS13,RSS09,RSS12}, $^{165}$Ho~\cite{TNM02}, $^{198}$Pt~\cite{SND13,SNL09}, $^{154}$Sm~\cite{GZH15}, $^{90}$Zr~\cite{KJP12}, $^{124}$Sn~\cite{PSP18}, and $^{197}$Au~\cite{PTN14} targets, and $^9$Be projectile on 
$^{89}$Y~\cite{PSC10}, $^{124}$Sn~\cite{PPS10}, $^{144}$Sm~\cite{GPC06a,GPC06b}, $^{186}$W~\cite{FGL13}, $^{169}$Tm~\cite{FGL15}, 
$^{187}$Re~\cite{FGL15}, $^{181}$Ta \cite{CFG14}, and $^{209}$Bi~\cite{DHS10} targets. Very recently, 
Cook  {\it et al.}~\cite{CSB19} performed an experiment to investigate the origin of the suppression of CF in collisions of $^7$Li. 

\medskip

The first CF and ICF  calculations in collisions of weakly bound nuclei were based on classical mechanics~\cite{HDH04,DHH02,DHT07,Dia10,Dia11}.
Although they were able to reproduce the main trends of the data, they missed important features of the fusion process, like tunnelling effects.  
Such effects were  approximately taken into account in semiclassical~\cite{MCD14,KCD18} models, which, however, were not fully satisfactory. 
More realistic quantum mechanical calculations, based on the continuum discretized coupled channel (CDCC) method, have also been developed. 
However, most calculations provided only the TF cross section~\cite{KKR01,DTB03,JPK14,DDC15}. The exceptions are the works of 
Hagino {\it al.}~\cite{HVD00} and Diaz-Torres and Thompson~\cite{DiT02}, which evaluated CF and TF in collisions of $^{11}$Be ($= ^{10}$Be + $n$) 
on  $^{208}{\rm Pb}$. In their method, the TF cross section was written as a sum of radial integral of the imaginary potential, in the subespace of bound
channels and in the subespace of {\it bins} (continuum discretized states). The former and the latter were then associated with CF and ICF, respectively. 
Since their imaginary potential depended exclusively on the projectile-target distance, cross terms between bound channels and bins  
vanished identically. This method is nice but it can only be used for projectiles like $^{11}$Be, which break up into a heavy 
charged fragment and a light uncharged one. It relies on the assumption that the center of mass of the projectile is close to that of the heavy fragment,
and far from the light one. Thus, it cannot be used for projectiles like $^7$Li, that breaks up into two fragments of comparable 
masses. In Ref.~\cite{PJK16} V. V. Parkar \textit{et. al.} proposed an approximation for the calculation of CF, ICF, and TF. The method consisted of 
performing separate CDCC calculations including different fragment - target short-range and projectile-target potentials to determine the cross-sections. 
Although the method has some success in describing the individual experimental cross-sections for the $^{6,7}$Li +$^{209}$Bi, $^{198}$Pt fusion 
reactions, it has the inconsistency that not all the fusion cross sections are determined from the same CDCC calculation.Very recently, Lei and 
Moro~\cite{LeM19} implemented a quantum mechanical approach based on the spectator-participant inclusive breakup 
model of Ichimura, Austern and Vincent~\cite{IAV85}. In their work, the CF cross section was indirectly determined subtracting from the total reaction 
cross section the contributions from inelastic scattering, elastic breakup and inclusive non-elastic breakup. The method was applied to the 
$^{6,7}{\rm Li}+^{209}{\rm Bi}$ systems and the results were shown to describe very well the experimental CF cross sections above the Coulomb 
barrier. Although the work of Lei and Moro are a significant advance on the existing theories, there is considerable room for improvements. One important 
point is that in this approach the ICF cross section is contained in the inclusive non-elastic breakup cross section. Thus, it cannot be evaluated individually.
Boseli and Diaz-Torres~\cite{BoD15} proposed a quantum-mechanical model to describe the time-evolution of wave packets. The method was used to 
study collisions of tightly bound systems~\cite{VoD19,DiW18} and also to estimate CF and ICF cross sections for the weakly bound $^7{\rm Li}+^{209}{\rm Bi}$
system. The method is promising but so far it has not been used in realistic calculations involving weakly bound projectiles.

\medskip

In this letter, we propose a new quantum mechanical method to derive CF and ICF (and hence TF) cross sections in collisions of any weakly bound projectile. 
The method is applied to the $^{7}{\rm Li}+^{209}{\rm Bi}$ system and the results are compared to the data of Dasgupta {\it et al.}~\cite{DHH02,DGH04}.\\

We consider the collision of a weakly bound projectile formed by two fragments, $c_1$ and $c_2$, on a spherical target.  The projectile-target
relative vector and the vector between the two fragments of the projectile are denoted by ${\bf R}$ and ${\bf r}$, respectively. The fragments
interact with the target though the complex potentials, $\mathbb{V}^{(i)}(r_{\scr i}) \equiv \mathbb{U}^{(i)}(r_{\scr i}) - i\,\mathbb{W}^{(i)}(r_{\scr i})$,
with $i=1,2$, where $r_i$ is the distance from the centers of fragment $c_i$ and the target. As in Refs.~\cite{HVD00,DiT02}, we start from the expression 
for the TF cross section within the CDCC approach,
\begin{eqnarray} 
\sigma_{\scr TF} &=& \frac{1}{|A|^2}\ \frac{k}{E}\ 
\  \left\langle\ {\rm \Psi}^{\scr (+)}\, \left| \, {\mathbb W}\,  \right| {\rm \Psi}^{\scr (+)}\  \right\rangle \nonumber\\
                            &=& \frac{1}{|A|^2}\ \frac{k}{E}\ \sum_{\alpha,\alpha^\prime} \
 \left\langle \psi_\alpha \left|\,  W_{\alpha,\alpha^\prime}\,
 \right|   \psi_{\alpha^\prime} \right\rangle,
\label{TF-1}
\end{eqnarray}
where the indices $\alpha$ and $\alpha^\prime$ run over bound channels and bins. Above, $\psi_\alpha({\bf R})$ is the projectile-target relative wave function in 
channel $\alpha$, and $W_{\alpha,\alpha^\prime}({\bf R})$ is the matrix-elements of the imaginary potential between channels $\alpha$ and $\alpha^\prime$.

Refs.~\cite{HVD00,DiT02} adopted a short-range imaginary potential, $W(R)$, that acts on the center of mass of the projectile, independently of it being bound
or unbound. Since it does not depend on ${\bf r}$, its matrix-elements are diagonal in channel space. In this way, the TF cross section of Eq.~(\ref{TF-1}) 
reduces to a sum of independent contributions from the channel. Hagino {\it et al.}~\cite{HVD00} then assigned the contributions from bound channels to CF, and
those of unbound channels to ICF.  However, their choice of the imaginary potential does not allow
 for the individual absorption of one of the fragments. To avoid this problem, we adopt the imaginary potential $\mathbb{W}({\bf R},{\bf r}) =\mathbb{W}^{(1)}(r_{\scr 1}) + 
 \mathbb{W}^{(2)}(r_{\scr 2})$, which is not diagonal in channel space. Then we keep off-diagonal matrix-elements within the same subespace ( bound (B) or 
 continuum (C)\,), but neglect off-diagonal matrix-elements between channels in different subespaces. The sum of Eq.~(\ref{TF-1}) then splits as,
$\sigma_{\scr TF} = \sigma^{\scr B}_{\scr TF}+ \sigma^{\scr C}_{\scr TF}$. As in Refs.~\cite{HVD00,DiT02}, we make the assumption: 
\begin{equation}
\sigma_{\scr DCF} = \sigma_{\scr TF}^{\scr B}.
\label{DCF}
\end{equation}

\bigskip

The physical meaning of $ \sigma_{\scr TF}^{\scr C}$ requires further discussion. Performing proper angular momentum expansions of the
wave functions and the imaginary potentials, $\sigma_{\scr TF}^{\scr C}$ can be put in the form
\begin{equation}
\sigma_{\scr TF}^{\scr C}= \frac{\pi}{k^2}\,\sum_J (2J+1)\ \left[ \mathcal{P}^{\scr (1)}(J) +  \mathcal{P}^{\scr (2)}(J) \right].
\end{equation}
Above, $\mathcal{P}^{\sc (i)}(J)$ is the probability of absorption of fragment $c_i$ in a collision with total angular momentum $J$. It corresponds 
to the contribution of the potential $\mathbb{W}^{\scr (i)}$ to the $J$-expanded version of Eq.~(\ref{TF-1}), with $\alpha$ and $\alpha^\prime$ 
constrained to channel in the continuum.\\

The ICF and the SCF cross sections can be determined using the probabilities $\mathcal{P}^{\scr (1)}(J)$ and $\mathcal{P}^{\scr (2)}(J)$. We follow 
the procedure of the semiclassical calculations of Ref.~\cite{KCD18}, writing for the ICF probabilities for fragments $c_1$ (ICF1) and $c_2$ (ICF2) as,

\begin{eqnarray}
\mathcal{P}^{\scr ICF1}(J) &=& \mathcal{P}^{\scr (1)}(J) \times \left[ 1\ -\ \mathcal{P}^{\scr (2)}(J)\right] ,\\
\mathcal{P}^{\scr ICF2}(J) &=& \mathcal{P}^{\scr (2)}(J) \times \left[ 1\ -\ \mathcal{P}^{\scr (1)}(J)\right] ,
\end{eqnarray}
The ICF cross sections of fragment $c_i$ is then given by
\begin{equation}
\sigma_{\scr ICFi} = \frac{\pi}{k^2}\,\sum_J (2J+1)\ \mathcal{P}^{\scr ICFi}(J) ,
\end{equation}
and the total ICF cross section is: $\sigma_{\scr ICF} = \sigma_{\scr ICF1} + \sigma_{\scr ICF2}$.

\smallskip 

Finally, the SCF cross section can be obtained subtracting $\sigma_{\scr ICF}$ from $\sigma_{\scr TF}^{\scr C}$. One gets
\begin{equation}
\sigma_{\scr SCF} = \frac{\pi}{k^2}\,\sum_J (2J+1)\ \Big[  2\ \mathcal{P}^{\scr (1)}(J) \times \mathcal{P}^{\scr (2)}(J)  \Big] .
\label{SCF}
\end{equation}
Then, the CF cross section is $\sigma_{\scr CF} = \sigma_{\scr DCF}+\sigma_{\scr SCF}$, 
with the DCF and SCF cross sections given respectively by Eqs.~(\ref{DCF}) and (\ref{SCF}).

\bigskip

Our method was used to evaluate CF and TF cross sections for the $^7$Li + $^{209}$Bi system, which have been measured by the ANL group~\cite{DHH02,DGH04} at 
near-barrier energies. For this purpose, we wrote the CF-ICF computer code (unpublished), based on the angular momentum projected version of the above equations,
using intrinsic and radial wave functions extracted from the CDCC version of the FRESCO code~\cite{Tho88}. $^7$Li is a weakly bound projectile, which breaks up into 
$^3$H and $^4$He, with the breakup threshold of 2.47 MeV. \\

For the real parts of the interaction between the fragments and the target, $\mathbb{V}^{\scr (1)}(r_1)$ and $\mathbb{V}^{\scr (2)}(r_2)$, we used the S\~ao Paulo 
potential~\cite{CPH97,CCG02} (SPP). The Coulomb barrier, given by the maximum of the potential $V(R) = \left\langle \varphi_0\left| \mathbb{V}^{\scr (1)} 
+ \mathbb{V}^{\scr (2)} \right| \varphi_0 \right\rangle$, is $V_{\scr B} = 28.2$ MeV. Owing to the low binding energy of the fragments in $^7$Li, the barrier is 1.2 MeV lower 
than the one of the SPP for the $^7$Li + $^{209}$Bi system, where this property is ignored. For the imaginary parts, we adopted typical short-range strong absorption 
potentials. We took Woods-Saxon functions with the depths, radii and diffusenesses: $W_0 = 50$ MeV, $R_{\rm w} = 1.0\,\left[ A_i^{\scr 1/3} +
A_{\scr T}^{\scr 1/3} \right]$ fm, where $A_i$ and $A_{\scr T}$ are respectively the mass numbers of $c_i$ and the target, and $a_{\scr w} =0.2$ fm. The continuum 
discretization was carried out with bins of orbital angular momenta up to $l_{\scr max} = 4\,\hbar$ and energies up to $\varepsilon_{\scr max} = 8$ MeV. For the $l=0,1,2$ 
and 4$\hbar$ we used bins of constant width, with density of 1 bin/MeV. At $l = 3\,\hbar$, we modified the mesh to account for the resonances at 2.2 and 4.2 MeV.
Around these energies the bins were much sharper, with densities up to 10 bins/MeV. We made sure that the continuum discretization with these parameters guaranteed 
good convergence in our calculations. \\

\begin{figure}
\begin{center}
\includegraphics*[width=8 cm]{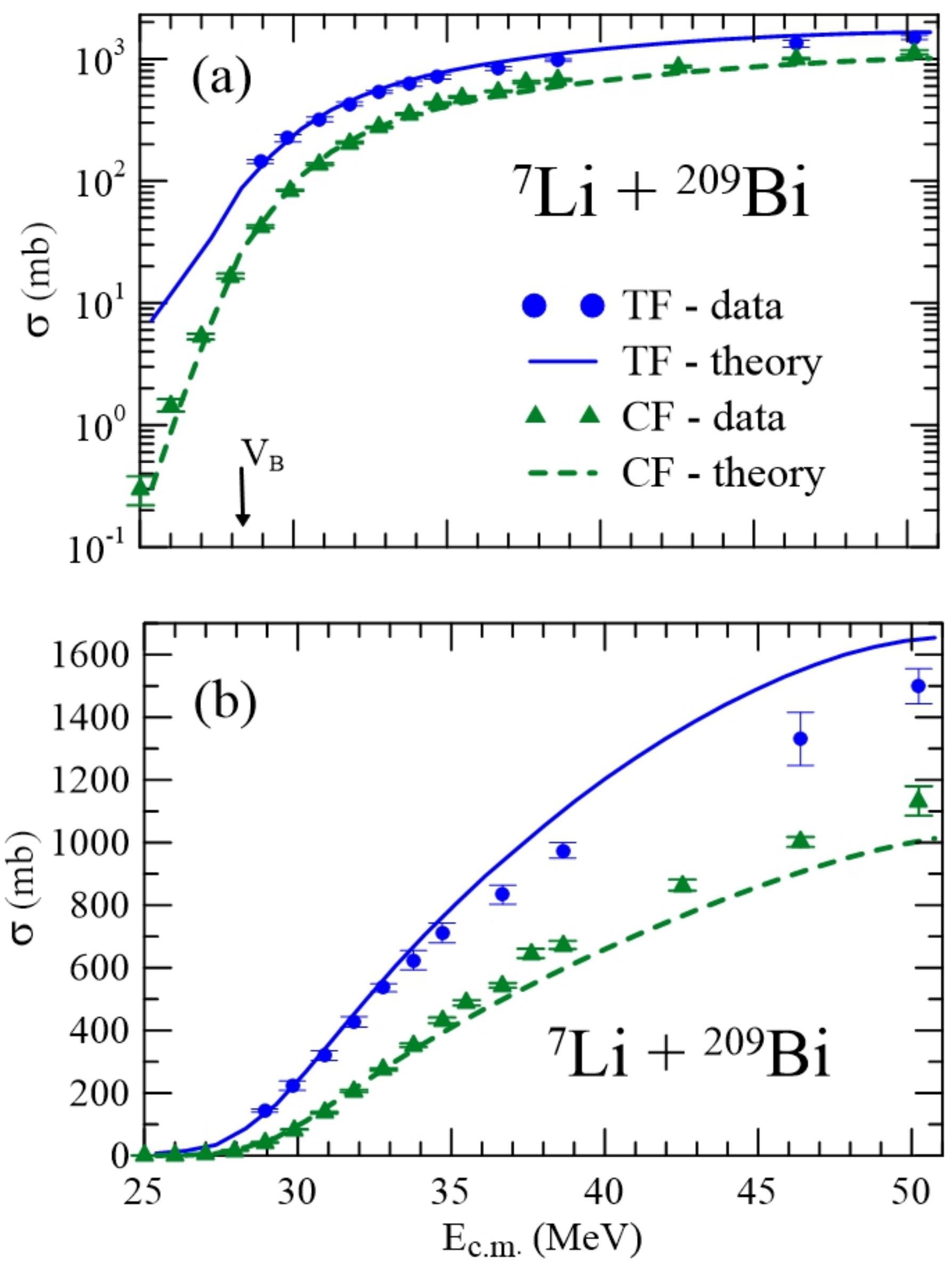}
\end{center}
\caption{(Color on line) Calculated TF and CF cross sections for the $^7$Li + $^{209}$Bi system at near-barrier energies, in comparison with the data
of Refs.~\cite{DHH02,DGH04} (blue solid circles and green triangles). The results are shown in logarithmic (panel (a)) and linear (panel (b)) scales.}
\label{Li7-Bi209}
\end{figure}
Fig.~\ref{Li7-Bi209} shows the TF and the CF cross sections predicted by our method in comparison with the data of Refs.~\cite{DHH02,DGH04}. At sub-barrier energies 
and above-barrier energies up to $E_{\rm c.m.} \sim 35$ MeV, the agreement between the calculated CF cross section and the data is excellent. Above 35 MeV, the agreement
is slightly poorer. Our calculations underestimate the  experimental CF cross section by $\sim 10\%$.\\

\begin{figure}
\begin{center}
\includegraphics*[width=8 cm]{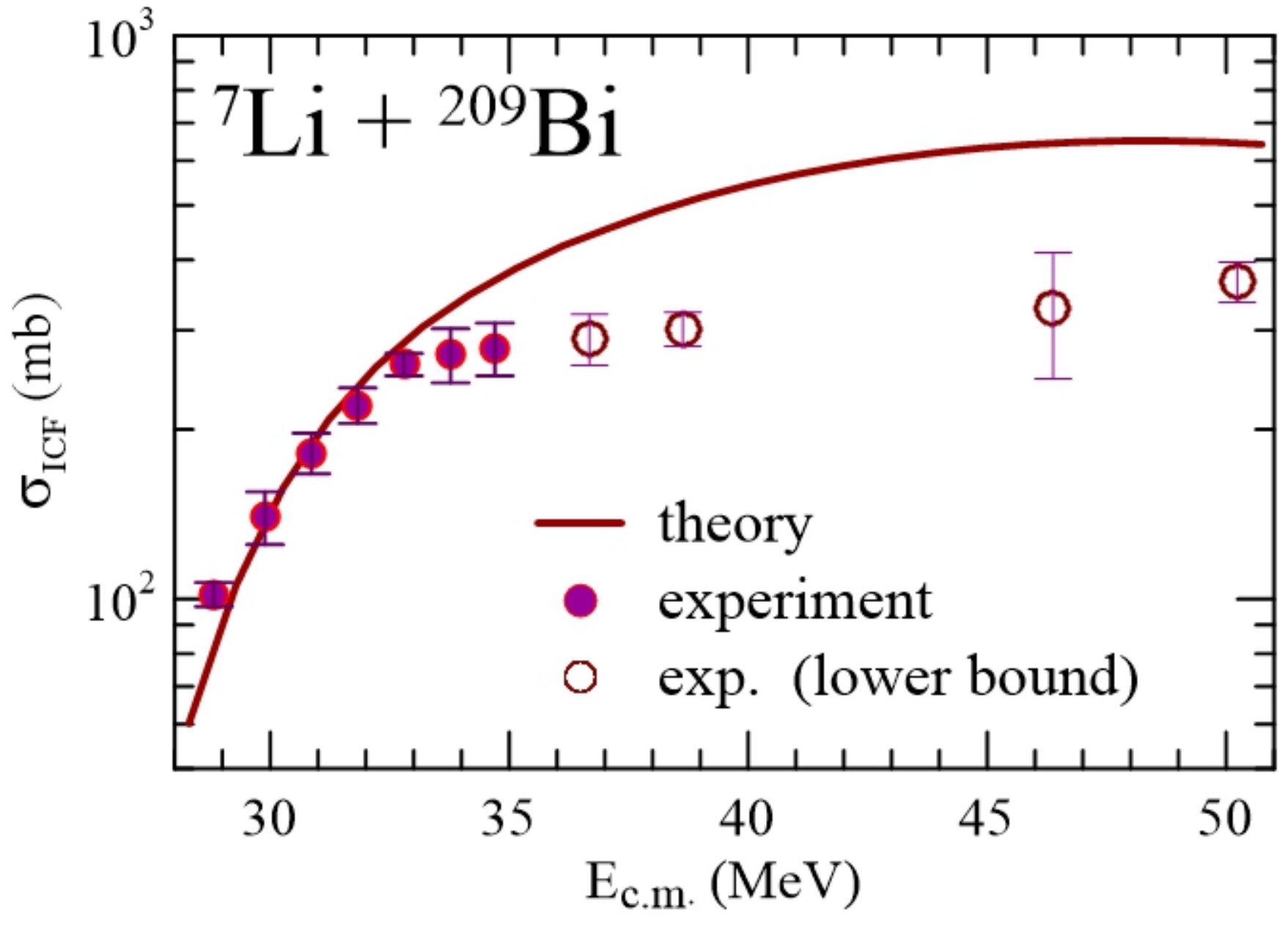}
\end{center}
\caption{(Color on line) Calculated ICF cross section  for the $^7$Li + $^{209}$Bi system, in comparison with the data of Refs.~\cite{DHH02,DGH04}.  Red 
solid circles are data points where all relevant decay channels were measured, whereas red open circles correspond lower bound to the cross section. For details, see the text. }
\label{ICF_T-E}
\end{figure}
The agreement between the theoretical TF cross section and the data below 35 MeV is also excellent. Above this energy, the theoretical cross section exceeds the experimental
results by up to $\sim 15\%$. In fact, the agreement in this case may be better.  The authors of the experiment pointed out that their ICF data above 36 MeV 
should be considered a lower bound to the actual cross section. The reason is that the contribution from $^{209}$Po, which is in the decay chain of the compound nuclei 
produced in both ICF processes ($^{213}$At and $^{212}$Po), cannot be measured owing to its long half life ($t_{\scr 1/2} = 102$ y), and estimates made with the code 
PACE~\cite{Gav80} indicate that
this channel becomes important above $\sim 36$ MeV. Since $\sigma_{\scr TF}$ contains $\sigma_{\scr ICF}$, the calculated TF cross section is consistent with the data. It is interesting
to compare also the ICF cross section predicted by our method and the data. This is done in Fig.~\ref{ICF_T-E}.  Clearly, the agreement between theory and experiment is excellent
up to $\sim 33-34$ MeV, and at higher energies the calculated cross section is larger than the data.\\
\begin{figure}
\begin{center}
\includegraphics*[width=8 cm]{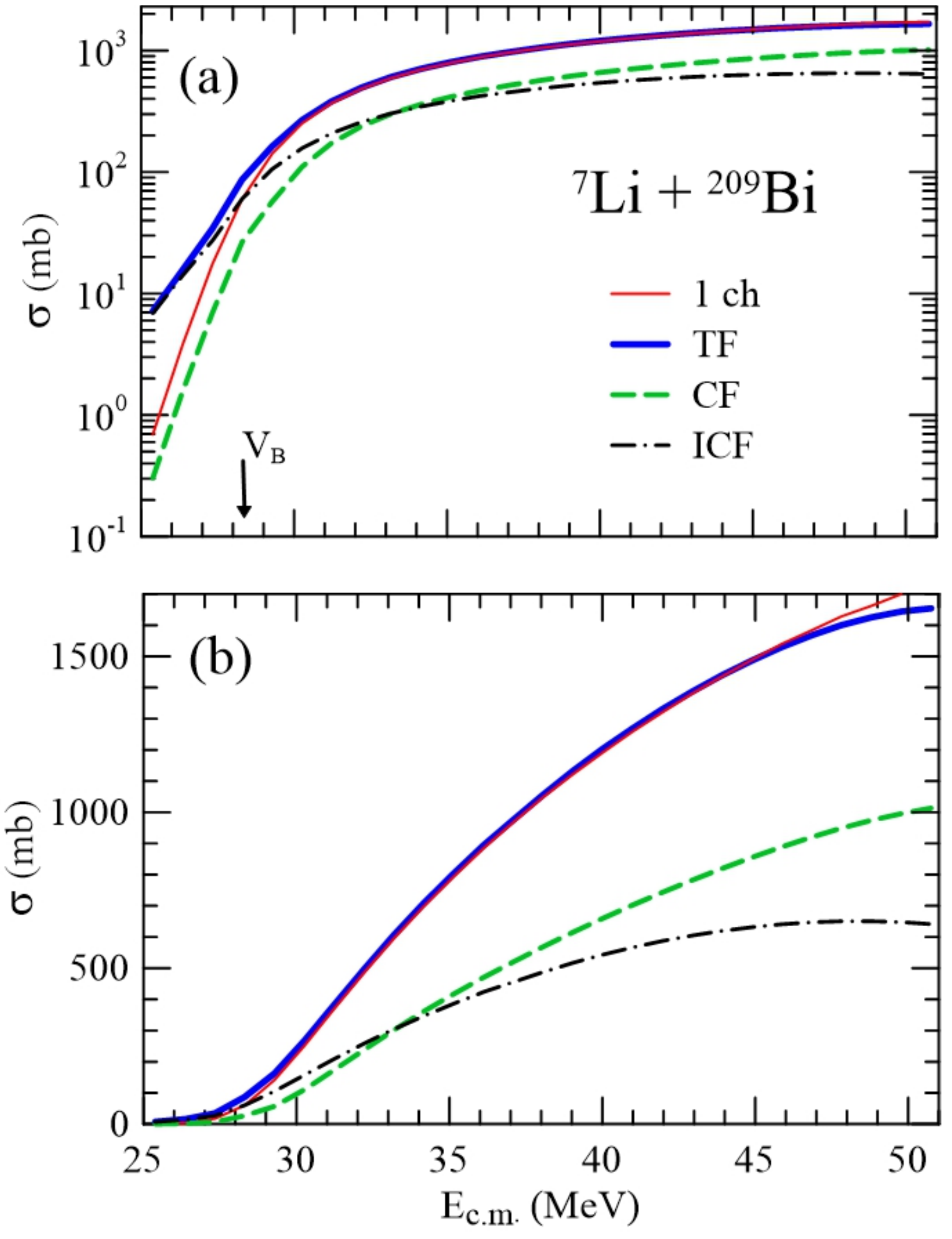}
\end{center}
\caption{(Color on line) Calculated CF, ICF and TF cross sections for the $^7$Li + $^{209}$Bi system, in comparison with the fusion cross section of 
a one-channel calculation, where couplings with the breakup channel are switched off. The results are show in logarithmic (panel (a)) and 
linear (panel (b)) scales. The height of the Coulomb barrier is indicated by $V_{\scr B}$ on panel (a).}
\label{CDCC-vs-1ch}
\end{figure}

The above comparison between theory and experiment indicates that the data of Refs.~\cite{DHH02,DGH04} can be very well described by our CDCC-based model, 
which considers only inelastic channels and breakup. At first sight, it seems to disagree with Ref.~\cite{CSB19}, which concluded that the main process contributing 
to the ICF data of Refs.~\cite{DHH02,DGH04} is direct triton stripping. But, in fact, we do not think there is any contradiction. We made coupled reaction channel 
calculations to estimate the triton-stripping cross section to bound states in $^{212}$Po and obtained cross sections orders of magnitude lower than the data. Thus, 
the stripping processes considered in Ref.~\cite{CSB19} must be mainly to the continuum of $^{212}$Po and, as has been pointed out in Ref.~\cite{PNT15}, breakup 
can be though of as transfer to the continuum of the target. \\

The influence of breakup couplings on fusion can be better understood comparing the CF, ICF and TF cross sections of our CF-ICF calculation with the one obtained with a 
one-channel calculation, where all couplings are switched off. This is done in Fig.~\ref{CDCC-vs-1ch}.  The behavior of the cross sections above the Coulomb barrier
can be observed more clearly in the linear plot of panel (b). The CF cross section is strongly suppressed with respect to the fusion cross section of the one-channel calculation. On the other hand, the TF cross section can hardly be distinguished from the cross section of the one-channel calculation. 
This indicates that breakup couplings split 
the one-channel cross section into CF and ICF components, keeping their sum practically unchanged. The behavior of the cross sections at sub-barrier energies can
be observed in the logarithmic plot of panel (a). There is still some suppression of CF, but it decreases with the collision energy. On the other hand, the TF cross section
below the Coulomb barrier is strongly enhanced. One sees that in this energy range the TF cross section is dominated by the ICF process. This is not surprising since 
at low energies the transmission coefficients for the lighter fragments must be much larger than that for the whole projectile. \\

\bigskip

We have proposed a new method to evaluate CF and ICF cross sections in collisions of weakly bound projectiles. Our method has the
advantages of being based on full quantum mechanics and being applicable to any weakly bound projectile that breaks up into two fragments. As an example, the 
method was used to calculate CF, ICF and TF cross sections in the $^7$Li + $^{209}$Bi collision, and the results were compared with experimental data. Considering that our 
calculations are based on standard heavy ion potentials and that it contains no free parameter, the overall agreement between theory and experiment is extremely good,
above and below the Coulomb barrier.\\

Although in its present stage our method is restricted to projectiles composed of two fragments, it can be generalized to projectiles that break up into three fragments,
like $^9$Be.\\

\section*{Acknowledgments}

Work supported in part by the Brazilian funding agencies, CNPq, FAPERJ, and the INCT-FNA (Instituto Nacional de Ci\^encia e 
Tecnologia- F\'\i sica Nuclear e Aplica\c c\~oes), research project 464898/2014-5. We are indebted to Dr. Jonas Ferreira for providing us with the CRC 
calculations of $t$-stripping, and to professor Raul Donangelo for critically reading the manuscript.

\bibliographystyle{apsrev}

\end{document}